\documentstyle[prb,aps,psfig,epsf,epsfig,twocolumn]{revtex}
\newcommand {\be}{\begin{equation}}
\newcommand {\ee}{\end{equation}}

\tighten

\begin{document}

\title{\bf Two-dimensional oriented self-avoiding walks with parallel contacts}
\author{G.T. Barkema, U. Bastolla, and P. Grassberger}
\address{HLRZ, Forschungszentrum J\"ulich, D-52425 J\"ulich, Germany}
\date{July 30, 1997}

\maketitle

\begin{abstract}
Two closely related models of oriented self-avoiding walks (OSAWs) on a
square lattice are studied.  We use the pruned-enriched Rosenbluth
method to determine numerically the phase diagram. Both models have
three phases: a tight-spiral phase in which the binding of parallel steps
dominates, a collapsed phase when the binding of anti-parallel steps
dominates, and a free (open coil) phase. We show that the system features a
first-order phase transition from the free phase to the tight-spiral
phase, while both other transitions are continuous.
The location of the phases is determined accurately. We also study turning
numbers and gamma exponents in various regions of the phase diagram.
\end{abstract}

\section{Introduction}
Many aspects of the behavior of polymers can be described by
self-avoiding walks on a lattice. To incorporate the interactions of
the polymer with itself, a binding energy $\epsilon$ can be assigned
if either two nearest-neighbour lattice sites are visited by the polymer
(point-contact model), or if two steps of the SAW are located on 
opposite sides of a plaquette of the lattice (step-contact model).
Both models show the same qualitative behaviour.

To describe the coil-globule (``theta'') transition, it is sufficient
to assume these interactions to be isotropic, but some polymers have
interactions that depend on the spatial orientation of the polymer, for
instance A-B polyester.  Such polymers are conveniently modeled by {\it
oriented} self-avoiding walks (OSAW) with short-ranged interaction
between steps depending on their relative orientation
\cite{cardy94,debbieetal95,flesia95,koo95,barkema96,prellberg97,trovato97}.
This orientation-dependent strength of the interaction can be
incorporated in the model by distinguishing {\it parallel} and {\it
anti-parallel} step-contacts, and assigning an {\it additional} binding
energy $\epsilon_p$ only between parallel step-contacts; with
point-contact energies, incorporating orientation-dependence in an
elegant manner is complicated by the ends of the polymer.

Most research on SAWs has been based on the point-contact model, since it
is numerically better behaved than the step-contact model.
Most research on OSAWs however has been based on the step-contact model,
in which the inclusion of the additional binding of parallel step-contacts
is more natural to the model. We will study both models in this manuscript.

Bennett-Wood {\it et al} \cite{debbieetal95} enumerated all
configurations up to SAWs with a length of $n=29$ and ordered them
according to their number of parallel and anti-parallel step-contacts.
These results showed the existence of three phases: a free SAW phase, a
normal collapsed phase and a compact spiral phase. The transition from
the free to the spiral phase was conjectured to be of first order.

For the case $\epsilon=0$ (only binding energies between parallel
step-contacts), Barkema and Flesia \cite{barkema96} extended
the exact enumeration of the OSAWs to length $n=34$. 
In the same paper, the energy of the ground state (for positive
$\epsilon_p$) and its degeneracy as a function of length was given, and an
approximation to the number of configurations $C_n(m_p)$ of polymers
of length $n$ with $m_p$ parallel step-contacts was proposed:
\begin{eqnarray}
\label{partfunc}
\nonumber
C_n(1) & = &  p_n C_n(0) \\
C_n(m) & \approx & C_n(1) \cdot \exp (-q_n (m-1)) \quad (m>1),
\end{eqnarray}
where $p_n$ and $q_n$ are $n-$dependent parameters.  The partition
function for $\epsilon=0$ can be constructed, and from this it was
concluded that the transition from the free phase to the spiral phase
at $\epsilon=0$ occurs at $\epsilon_p=\log(\mu)=0.9701$ (where 
$\mu =2.638$ is the growth constant for SAWs\cite{conway93}) and 
is a first order transition. We use units such that $k_BT=1$.

For the step-contact model, Prellberg and Drossel \cite{prellberg97}
argued that the transition from the collapsed phase to the spiral phase
is continuous.  They also argued that the location of the
theta-transition (from the free to the collapsed phase) is independent
of $\epsilon_p$.

Trovato and Seno \cite{trovato97} performed transfer matrix
calculations on the point-contact model. They also made some
calculations for the step-contact model, but with much less conclusive
results.  They found that the transition from either the free or the
collapsed to the spiral phase is probably of first order.

In this paper, we employ the pruned-enriched Rosenbluth method (PERM)
\cite{grassberger97,bastolla97,frauenkron97} to study the phase diagram 
of both models for two-dimensional OSAWs. The only deviation from 
the algorithm as described in the above references is that new steps
were biased both towards large numbers of contacts and large absolute
values of the turning number, with different biases in different parts 
of the phase diagram. As usual in PERM, this is corrected for
by reweighting.

The manuscript is organized as follows.
In section \ref{free2spiral} we study the phase transition from the
free phase towards the spiral phase in the step-contact model. To do
this, we determine numerically the partition function along three lines
$\epsilon=$constant, for all $\epsilon_p$.

In section \ref{col2spiral}, we use the same technique as in section
\ref{free2spiral} to study the transition from the collapsed to the
spiral phase, but for different values of $\epsilon$ which 
are above the collapse energy $\epsilon_{\theta}$.

The next section, section \ref{free2col}, studies the transitions for
$\epsilon$ close to the collapse point $\epsilon_{\theta}$.  To do this,
we use the point-contact model, since for this model finite-size
corrections at $\epsilon=\epsilon_{\theta}$ are smaller, and
simulations using PERM have smaller statistical fluctuations.

In section \ref{phasediagram}, we present our numerically determined
phase diagram for both models, discuss the nature and location of all
phase transitions, and summarize the results.

\section{Transition from the free phase to the spiral phase}
\label{free2spiral}

\begin{figure}
\epsfxsize=8cm
\epsfbox{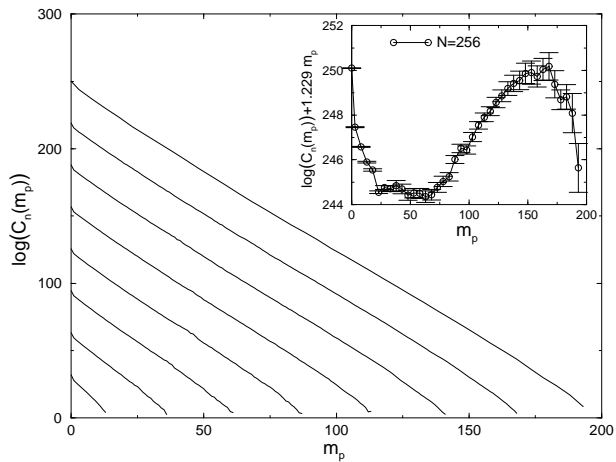}
\caption{Logarithm of the contribution to the partition function
$C_n(m_p)$ of configurations with $m_p$ parallel contacts, in the
absence of anti-parallel interactions (i.e., $\epsilon=0$). The lines
correspond to chain lengths of $n=32, 64, \dots, 256$ steps. The inset
shows the deviation from a straight line, by plotting
$\log(C_{256}(m_p))+1.229~m_p$ as a function of $m_p$. The presence of two
peaks indicates a first-order phase transition.
\label{qa=1}}
\end{figure}

With PERM, we measured $C_n(m_p)$, the contribution to the partition
function for walks of $n$ steps from configurations with $m_p$ parallel
contacts.  First, we did this at $\epsilon=0$, so that we can compare
our results with previous work.  In figure \ref{qa=1} we have plotted
$\log(C_n(m_p))$ as a function of $m_p$ for various chain lengths $n$.
To a good approximation, the curves are straight lines, in agreement
with the guess eq. (\ref{partfunc}).  The inset shows the deviation
from the straight line for $n=256$, by plotting $\log(C_n(m_p))+1.229~m_p$.
The figure combines data obtained at several values of $\epsilon_p$ close
to the transition.

{}From the partition function plot we conclude that there is a
first-order transition from the free phase, where configurations with
no parallel contacts dominate, to the spiral phase, where
configurations with many parallel contacts dominate. A closer look
shows that the curves in fig.\ref{qa=1} are not completely straight but
S-shaped, with a "bump" at some number $m_b^*$ of parallel contacts.
Looking at increasing lengths $n$ of the OSAWs, we see no evidence that
asymptotically $m_b^*/n \rightarrow 0$ nor $m_b^*/n \rightarrow 1$, and
therefore conclude that this bump will persist in the thermodynamic
limit ($n \rightarrow \infty$).  To extract the transition temperature
$\epsilon_{p,{\rm crit}}(n)$, we have determined for which $\epsilon_p$
the top of the bump and the walks with no parallel contacts contribute
equally to the partition function.  The results are $\epsilon_{p,{\rm
crit}}(n)$=2.11(3), 1.6(1), 1.42(4), 1.368(7), 1.315(2), 1.278(2),
1.250(2); and 1.229(2) for $n=32, 64, \dots, 256$ respectively.

Since the number of parallel contacts in a tight spiral scales as
$n-4\sqrt{n}$, we expect corrections of order $n^{-1/2}$ to the critical
temperature. Most likely, there are also corrections of order $n^{-1}$,
and possibly other corrections. Assuming however that corrections of order 
$n^{-1/2}$ are the leading ones, we extrapolated our values for
$\epsilon_{p,{\rm crit}}(n)$ to the limit $n \rightarrow \infty$, and
obtained for $\epsilon=0$ in the thermodynamic limit
$\epsilon_{p,{\rm crit}}=0.90(5)$, with the error mainly due to the
uncertainty in the finite-size correction.

We also studied the contribution to the partition function as a
function of the turning number $t$: the number of turns that the walk
has made clockwise, minus the number of turns anti-clockwise.  Note
that the turning number is not equal to the winding number $w$ 
as defined by Duplantier and Saleur \cite{dupla-saleur}; for
large chains in the free and the collapsed phase however, the two
quantities seem to be related by
$\langle (\frac{\pi}{2} t)^2 \rangle = 2 \langle w^2 \rangle$.
The factor of two is explained by observing that the turning number
receives contribution from both ends of the chain, while only one end
contributes to the winding number. For the spiral ground state, the turning
number is roughly equal to $2\sqrt(n)$.

\begin{figure}
\epsfxsize=8cm
\epsfbox{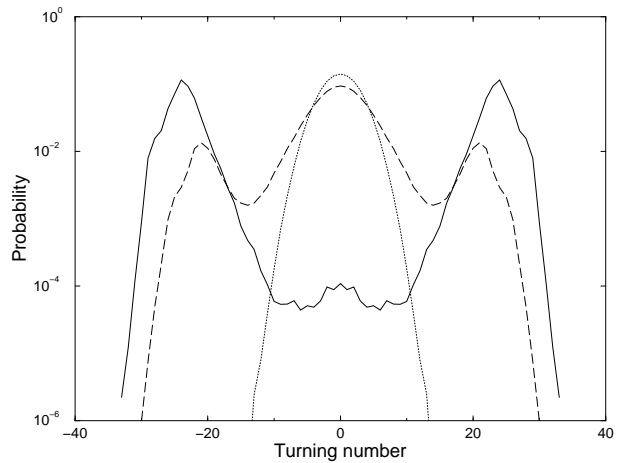}
\caption{Probability of turning number $t$ as a function of
$t$, for chains of length $n=256$ for the case $\epsilon=0$.
Different curves correspond to $\epsilon_p=0$ (dotted line),
$\epsilon_p=1.194$ (dashed line), and $\epsilon_p=1.253$ (solid line).
\label{turn1}}
\end{figure}

Results for $n=256$ are presented in figure \ref{turn1}, where we
averaged the histograms for positive and negative turning numbers.
We observe that for $\epsilon_p=1.253$, the histogram has two peaks
at $\pm 23$: we are in the spiral phase. At $\epsilon_p=0$, the turning
number is between -10 and 10, with a maximum at zero: we are outside
the spiral phase. At $\epsilon=1.194$ we are close to the
transition, and the peaks at $t=\pm 21$ and the peak around zero
coexist. The fact that in the histogram for turning numbers the peaks
maintain their location, while their relative importance changes, is
consistent with our earlier conclusion that the phase transition is
first-order. For a continuous transition, we would expect that the two
peaks in the spiral phase would approach zero gradually.

We repeated this procedure for $\epsilon=0.993$ and for the case where
anti-parallel contacts are strictly forbidden ($\epsilon \rightarrow -\infty$)
but parallel contacts have a finite binding energy ($\epsilon+\epsilon_p$
is finite).  Qualitatively, the behavior is the same as for
$\epsilon=0$, and we conclude that the transition is also first-order.
The bump seems to shift to the left with increasing $\epsilon$.
For $\epsilon=0.993$ we obtained in the thermodynamic limit
$\epsilon_{p,{\rm crit}}=0.05(5)$, while if anti-parallel contacts are
strictly forbidden, we find $\epsilon_{p,{\rm crit}}+\epsilon=0.75(5)$.

\section{Transition from the collapsed phase to the spiral phase}
\label{col2spiral}

To study the transition from the collapsed phase to the spiral phase in
the step-contact model,
we used the same technique as in section \ref{free2spiral}: at a
particular value for $\epsilon$, we calculated the contribution of
configurations to the partition function as a function of the number of
its parallel contacts. This allows us to determine both the nature and
the location of the phase transition at the particular value of
$\epsilon$.

\begin{figure}
\epsfxsize=8cm
\epsfbox{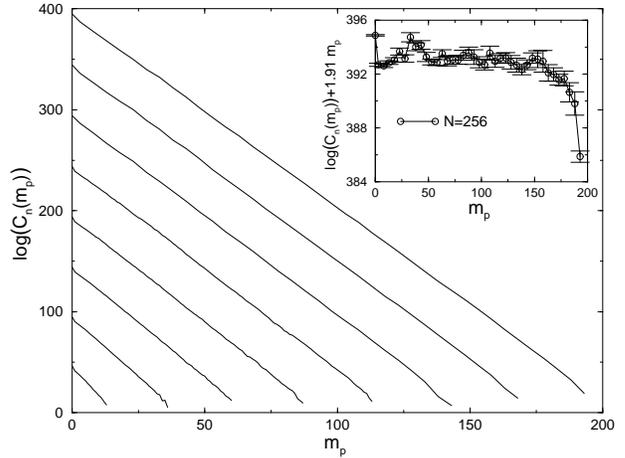}
\caption{
Logarithm of the contribution to the partition function $C_n(m_p)$ of
configurations with $m_p$ parallel contacts, for the case $\epsilon=1.609$
and no interaction between parallel steps ($\epsilon + \epsilon_p=0$).
Different curves correspond to chain lengths $n=32, 64, \dots, 256$
steps. The inset shows the deviation from the straight line by plotting
$\log(C_{256}(m_p))+1.91~m_p$ as a function of $m_p$. The two peaks that 
were present in the case $\epsilon=0$ have nearly disappeared and are 
compatible with being finite size effects, indicating
that the first-order phase transition has changed into a continuous one.
\label{qa=5b}}
\end{figure}

We used this method for $\epsilon=$1.253, 1.435 and 1.609.  These
values are well above the theta-transition, which is estimated to be
$\epsilon=1.21$ (see section \ref{free2col}).  Results analogous to
those shown in fig.\ref{qa=1}, but now for $\epsilon=1.609$, are
presented in figure \ref{qa=5b}. There are still two maxima in the
histogram -- one at $m_p=0$ and the other at $m_b^*>0$ -- but the
valley between them is very shallow and much more narrow.  Comparison
of different chain lengths now suggests that in the thermodynamic limit
$m_b^*/n$ approaches zero, indicating that the transition has become
continuous.

We obtained estimates for the location of the phase transition line
that are consistent with $\epsilon_{p,{\rm crit}}=0$.

\begin{figure}
\epsfxsize=8cm
\epsfbox{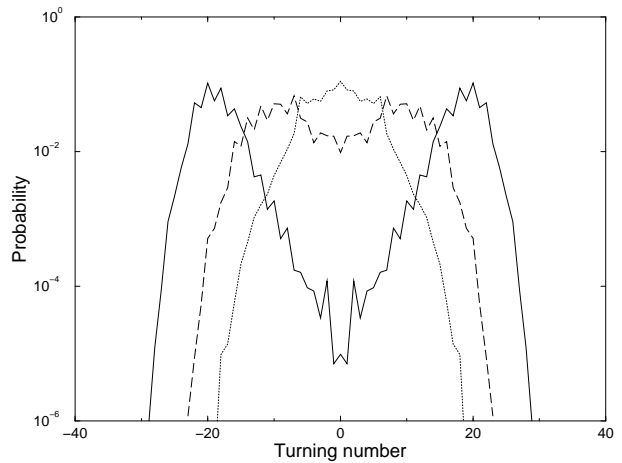}
\caption{Probability of turning number $t$ as a function of $t$, for
chains of length $n=256$, in the case $\epsilon=1.609$.  Different
curves correspond to $\epsilon_p=1.82$ (dotted line), 1.92 (dashed line),
and $\epsilon_p=1.96$ (solid line).
\label{turn5}}
\end{figure}

The histogram of the turning numbers is plotted in figure \ref{turn5}.
The picture is quite different from that of figure \ref{turn1}: the peaks
at high positive and negative turning numbers do not maintain their location
if the transition line is approached, but shift towards zero, where they
merge. This is consistent with a continuous phase transition.

\section{Transition from the free to the collapsed phase}
\label{free2col}

At first, we simulated the step-contact model at various values of
$\epsilon$, to obtain a precise value of $\epsilon_\theta$. Requiring
that the end-to-end distance scales as $n^{4/7}$ and the partition sum
as $\mu_\theta^n n^{1/7}$ (see Duplantier and Saleur~\cite{dupl-sal2}),
we got $\epsilon_{\theta}=1.21(2)$, independent of the value of
$\epsilon_p$, as long as $\epsilon_p<0$.  Next, we simulated the
ordinary (non-oriented) point-contact model at various values of
$\epsilon$, to obtain a precise value of $\epsilon_{\theta}$, and we
got $\epsilon_{\theta}=0.667(1)$, in good agreement with earlier estimates
\cite{meiro,hegg-grass}.

\begin{figure}
\begin{center}
\epsfxsize=8cm
\psfig{file=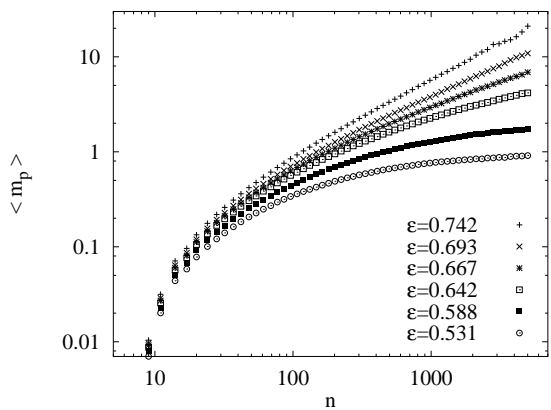,scale=.34,angle=270}
\end{center}
\caption{Log-log plot of the average number of parallel bonds versus 
chain length for different values of $\epsilon$, and for $\epsilon_p=0$. 
The $\theta$-point is at $\epsilon=0.667$. }
\label{fig5.ps}
\end{figure}

This difference in $\epsilon_{\theta}$ is due to the fact that point
contacts are roughly twice as frequent as bond contacts near the theta
point, and it makes simulations in the regime $\epsilon \geq
\epsilon_{\theta}$ much harder in the step-contact model than in the
point-contact model: due to the large value of $\epsilon$, the
Boltzmann weights of different configurations fluctuate strongly, which
creates problems for PERM. The point-contact model can be simulated
more efficiently by PERM (error bars decrease by roughly one order of
magnitude for the same CPU times), and systematic errors due to
finite-size corrections decrease, although they stay sizeable in both
models (the same was found by Trovato and Seno \cite{trovato97} for
transfer matrix calculations).

For this reason we used the point-contact model to study transitions
for $\epsilon\approx \epsilon_\theta$ in detail. This includes the
coil-globule transition for $\epsilon_p<0$ which happens exactly at
$\epsilon= \epsilon_\theta$, as well as the region around the triple
point $(\epsilon=\epsilon_\theta, \epsilon_p=0)$.

\begin{figure}
\begin{center}
\epsfxsize=8cm
\psfig{file=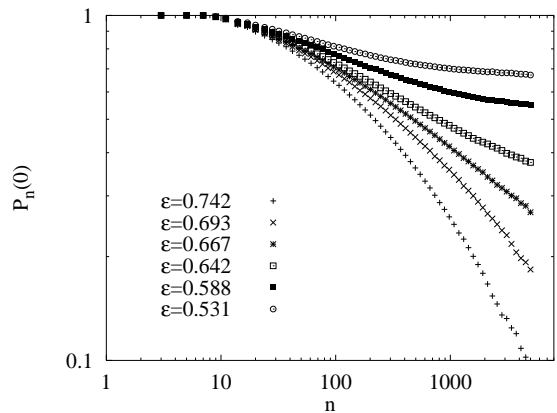,scale=.34,angle=270}
\end{center}
\caption{Log-log plot of $P_n(0)$, the chance to have no parallel bonds
in an $n$-step walk.}
\label{fig6.ps}
\end{figure}

In the same runs we also measured the average number of parallel
contacts. Results are shown in fig.\ref{fig5.ps}. They indicate that
$\langle m_p\rangle$ converges for $n\to\infty$ to a finite value, in
agreement with Barkema and Flesia \cite{barkema96}, as long as
$\epsilon<\epsilon_\theta$. This is however no longer true for
$\epsilon\geq \epsilon_\theta$. Exactly at the $\theta$-point, $\langle
m_p\rangle$ increases roughly as $\sqrt{n}$ for large $n$.  But finite
size corrections are so large that it is not clear whether this is
really the asymptotic behavior.  Thus, while parallel bonds are
unimportant below the $\theta$-point, they become important above it.
Consistent with this, we found that the probability $P_n(0) =
C_n(0)/\sum_m C_n(m)$ of having no parallel bond at all decreases to
zero for $\epsilon\geq \epsilon_\theta$, while it converges to a finite
value for $\epsilon<\epsilon_\theta$ (fig.\ref{fig6.ps}). Based on the
analogy with parts of percolation cluster hulls, it was conjectured by
Prellberg and Drossel \cite{prellberg97} that $P_n(0) = n^{-2/7}$ exactly at
the $\theta$-point. This is consistent with our data, although our data
show again very slow convergence and would suggest an exponent $\approx
1/4$ rather than 2/7.

The fact that $P_n(0)$ does not decrease exponentially with $n$ for
$\epsilon<\epsilon_\theta$ shows that indeed the open coil/globule
collapse occurs for all $\epsilon_p<0$ at the same value of $\epsilon$,
namely $\epsilon=\epsilon_\theta$
\cite{debbieetal95,trovato97,prellberg97}. Parallel contacts are simply
too rare to effect phase boundaries for $\epsilon_p<0$.  On the other
hand, the fact that $\langle m_p\rangle$ diverges at $\epsilon_\theta$
implies that this is no longer true for $\epsilon_p>0$. Thus the phase
boundary has a singularity at
$(\epsilon,\epsilon_p)=(\epsilon_\theta,0)$, which suggests that this
point is indeed the triple point where all three phase boundaries meet
\cite{trovato97}.

\begin{figure}
\begin{center}
\begin{minipage}[bh]{10.0cm}
\epsfig{file=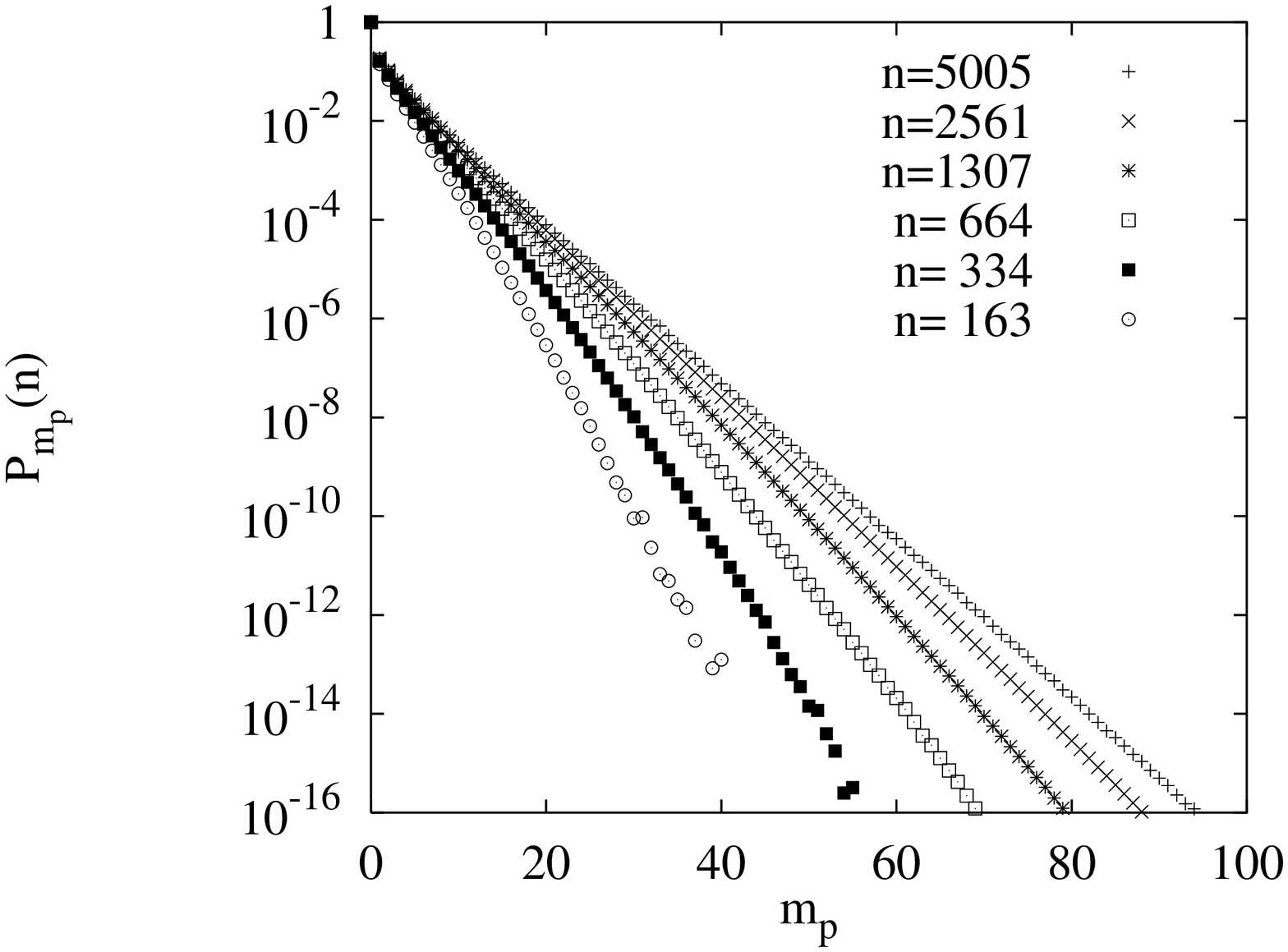,scale=.34,angle=270}
\epsfig{file=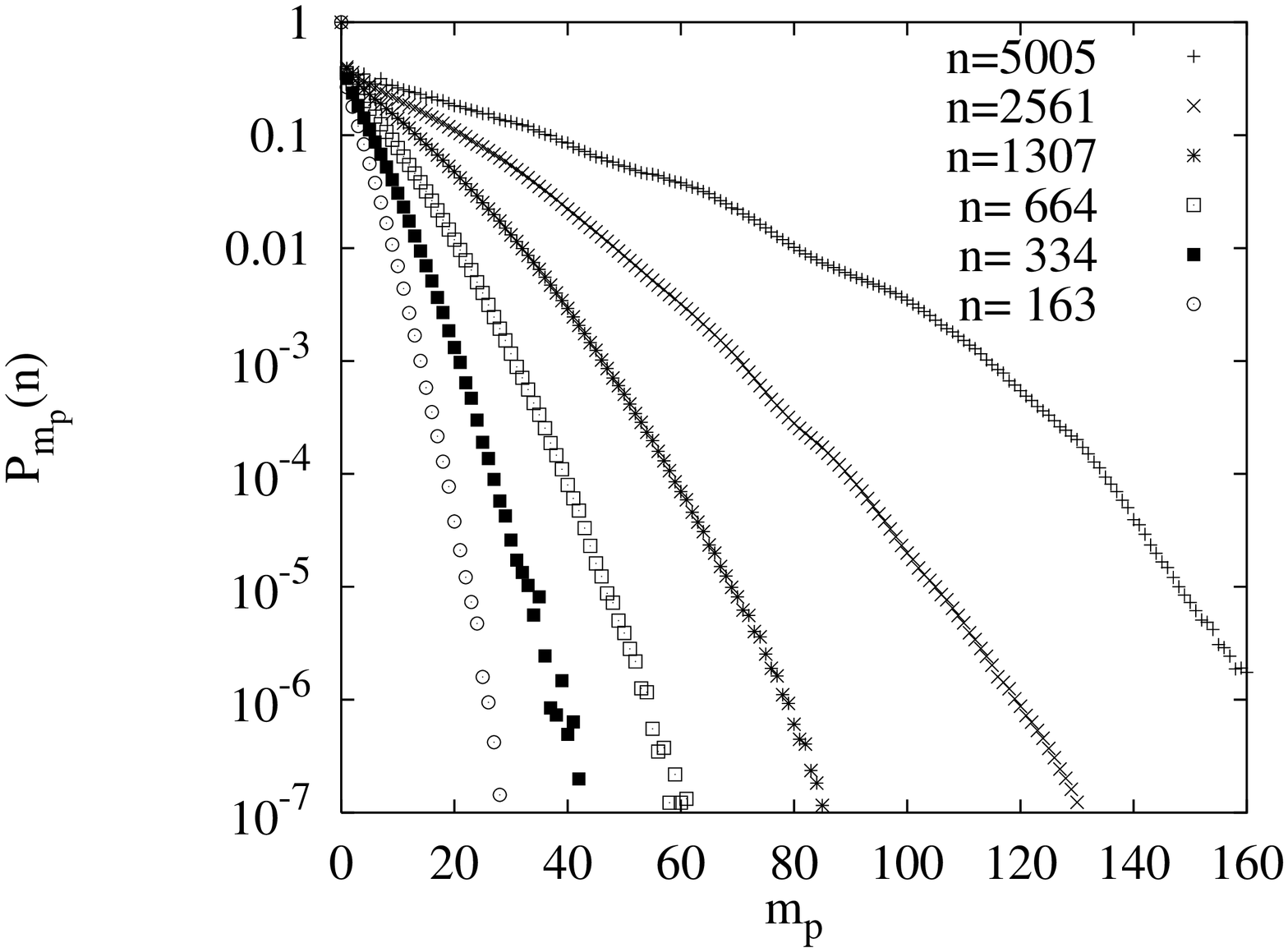,scale=.34,angle=270}
\end{minipage}
\end{center}
\caption{Distributions $P_n(m_p)$ for finding $m_p$ parallel bonds in 
chains of length $n$, normalized to $P_n(0)=1$. Each panel contains 
curves for $n=163, 334, 664, 1307, 2561,$ and 5005. Panel (a) is for 
$\epsilon = 0.531<\epsilon_\theta$, while panel (b) is for 
$\epsilon = 0.742>\epsilon_\theta$.}
\label{fig7.ps}
\end{figure}

In order to verify this and to determine the orders of the 
coil-spiral and globule-spiral transitions, we measured also
the distribution 
\be
   P_n(m_p) = C_n(m_p)/\sum_m C_n(m). 
\ee
Typical results 
for $\epsilon<\epsilon_\theta$ and for $\epsilon>\epsilon_\theta$ 
are shown in panels (a) and (b) of fig.\ref{fig7.ps}, respectively.
While $P_n(m_p)$ decreases roughly exponentially with $m_p$ in both 
plots, details are 
rather different. In panel (a) the exponent is nearly 
independent of $n$, suggesting that it is non-zero also for 
$n\to\infty$. Thus, the free - spiral transition happens at a positive 
$\epsilon_p$. In contrast, the exponent depends strongly on $n$ 
in panel(b), and seems to converge to zero for $n\to\infty$.
This is confirmed by a more careful analysis. It shows that the 
collapsed - spiral transition happens exactly at $\epsilon_p=0$, 
as suggested by Trovato and Seno \cite{trovato97}.

\begin{figure}
\begin{center}
\epsfxsize=8cm
\psfig{file=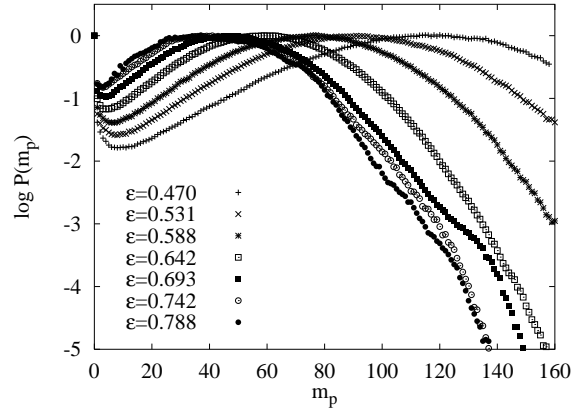,scale=.34,angle=270}
\end{center}
\caption{Log-log plot of $e^{a(\epsilon)m_p}P_n(0)$, with $a(\epsilon)$ 
such that both peaks have the same height. Again normalization is 
such that $P_n(0)=1$. }
\label{fig8.ps}
\end{figure}

To determine the order of the transitions, we plot 
$e^{a(\epsilon)m_p}P_n(m_p)$ with the parameter $a(\epsilon)$ determined 
such that both peaks in this function have the same height (compare 
the inserts in figs.\ref{qa=1} and \ref{qa=5b}). This is done 
for several values of $\epsilon$, but only for a single chain 
length ($n=2561$). Again the data are normalized to $P_n(0)=1$. 
Results are shown in fig.\ref{fig8.ps}. We see that there are two 
peaks for all values of $\epsilon$, but that the right peak is 
located at very small values of $m_p$ in the collapsed region, and 
moves to larger values of $m_p$ only if we go with $\epsilon$ below 
the $\theta$-point. Thus we see again that the double peak structure 
is a finite-size effect in the collapsed phase, as we had already 
seen in sec.\ref{col2spiral}, and that the collapse-to-spiral transition 
is second order.

\begin{figure}
\begin{center}
\epsfxsize=8cm
\psfig{file=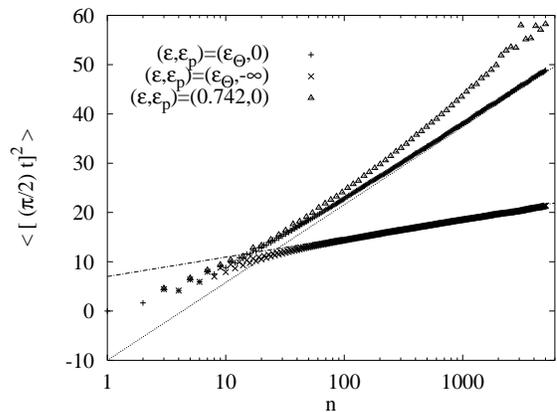,scale=.34,angle=270}
\end{center}
\caption{Average squared turning number $\langle t^2\rangle$ against 
$\log n$, for $(\epsilon,\epsilon_p)=(\epsilon_\theta,0)$ (+), 
$(\epsilon_\theta,-\infty)$ ($\times$),  and 
$(0.742,0)$ ($\triangle$).
The straight lines are the theoretical predictions for the first two 
cases. }
\label{fig9.ps}
\end{figure}

Finally, we also measured turning numbers at and near $\epsilon=
\epsilon_\theta$. Average squared turning numbers at the triple point 
$(\epsilon,\epsilon_p)=(\epsilon_\theta,0)$ and at the point 
$(\epsilon_\theta,-\infty)$ are shown in fig.\ref{fig9.ps}. Apart 
from the by now familiar large deviations for small $n$, we see clear 
indications for logarithmic laws
$\langle (\frac{\pi}{2} t)^2 \rangle = 2 \langle w^2 \rangle = 2C\log n$.
The constants $C$ are fully compatible with the predictions $C=24/7$ at
$(\epsilon_\theta,0)$ \cite{dupla-saleur} and 6/7 at
$(\epsilon_\theta,-\infty)$ \cite{prellberg97} which are indicated in
fig.\ref{fig9.ps} by straight lines. We have not measured turning
numbers with similar precision in other phases, but the overall picture
seems fully compatible with that of Prellberg and Drossel
\cite{prellberg97}.  On the collapsed-spiral transition line
$(\epsilon>\epsilon_\theta, \epsilon_p=0)$, $\langle t^2\rangle$ seems
to increase faster than $\log n$ (see also fig.\ref{fig9.ps}), but our
data are less precise there.

\section{Conclusions and discussion of the phase diagram}
\label{phasediagram}
We have found three phases for two-dimensional OSAWs: two of them, the
free phase and the collapsed phase, also exist in normal SAWs, and have
a turning number around zero; the third phase, the spiral phase, is
unique for OSAWs, and has a high turning number.

In section \ref{free2col}, we have confirmed that the location of
the transition from the free to the collapsed phase is
independent of the strength of parallel interactions. The critical value
is estimated to be $\epsilon_{\theta}=1.21(2)$ for the step-contact model,
and $\epsilon_{\theta}=0.667(1)$ for the point-contact model. This transition
also exists for SAWs, and is known to be continuous \cite{}.

The transition from the collapsed to the spiral phase is found to be
also continuous, and located at $\epsilon_p=0$, in agreement with 
theoretical predictions~\cite{prellberg97}.

\begin{figure}
\epsfxsize=8cm
\epsfbox{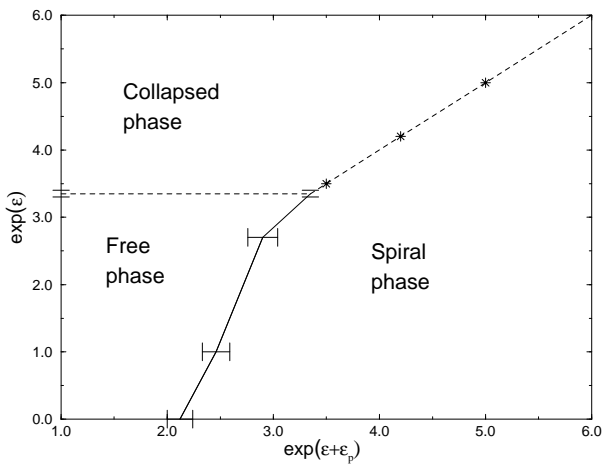}
\caption{Schematic drawing of the phase diagram of the step-contact model.
The model has three phases, a free phase, a collapsed phase, and a spiral
phase. The transitions from the collapsed phase to the free or the spiral
phase are continuous, the transition from the free to the spiral phase is
of first order. The transition line from the collapsed to the spiral phase
is located at $\epsilon_p=0$. The transition from the free to the collapsed
phase is located at $\epsilon=1.21(2)$. We determined three points on the 
transition line from the free to the spiral phase, and 
connected these points as a guide to the eye.
\label{phaseplot}}
\end{figure}

In section \ref{free2spiral}, we concluded that the transition from the
free phase to the spiral phase is a first-order one. For the step-contact
model, we have located three points $(\epsilon+\epsilon_p,\epsilon)$
on the phase transition line:  (0.75, $-\infty$); (0.90,0), and (1.04, 0.993).
The results for the step-contact model are combined in figure \ref{phaseplot},
where the phase diagram of the step-contact model is presented. The phase 
diagram for the point contact model is identical except for the detailed 
location of the transition lines.

The probability $P_n(0)$ of having no parallel bond at all decreases to
zero for $\epsilon\geq \epsilon_\theta$, while it converges to a finite
value for $\epsilon<\epsilon_\theta$.  Exactly at the theta-point, it
was conjectured by Prellberg and Drossel \cite{prellberg97}
that $P_n(0) = n^{-2/7}$. This is consistent with our data, although
slightly smaller exponents are not ruled out.

At the triple point $(\epsilon,\epsilon_p)=(\epsilon_\theta,0)$ and at
the point $(\epsilon_\theta,-\infty)$, the average squared turning numbers
grow logartihmically with $n$ with constants as predicted by Duplantier
and Saleur \cite{dupla-saleur}, and Prellberg and Drossel \cite{prellberg97}.

\section*{Acknowledgements}

GTB likes to thank S. Flesia for useful discussion. PG is supported by the 
Deutsche Forschungsgemeinschaft through SFB 237.


\begin{thebibliography}{10}
 
\bibitem{cardy94}
J.L.~Cardy,
\newblock Nucl. Phys. B. {\bf 419}, 411 (1994).

\bibitem{debbieetal95}
D.~Bennet-Wood, J.L.~Cardy, S.~Flesia, A.J.~Guttmann, and A.L.~Owczarek,
\newblock J. Phys. A {\bf 28}, 5143 (1995).

\bibitem{flesia95}
S.~Flesia, Europhys. Lett. {\bf 32}, 149-154 (1995).

\bibitem{koo95}
W.M.~Koo, J. Stat. Phys. {\bf 81}, 561 (1995).

\bibitem{barkema96}
G.T.~Barkema and S.~Flesia, J. Stat. Phys. {\bf 85}, 363 (1996).

\bibitem{prellberg97}
T.~Prellberg and B.~Drossel, cond-mat/9704100 (1997).

\bibitem{trovato97}
A.~Trovato and F.~Seno, Phys. Rev. E (1997).

\bibitem{conway93}
A.~Conway, I.G.~Enting and A.J.~Guttmann, J. Phys. A {\bf 26}, 1519 (1993).

\bibitem{grassberger97}
P.~Grassberger, to appear in Phys. Rev. E (1997).

\bibitem{bastolla97}
U.~Bastolla and P. Grassberger, preprint cond-\linebreak mat/9705178 (1997).

\bibitem{frauenkron97}
H. Frauenkron, U. Bastolla, E. Gerstner, P. Grassberger, and W. Nadler, 
preprint cond-mat/9705146 (1997).

\bibitem{dupla-saleur}
B. Duplantier and H. Saleur, Phys. Rev. Lett. {\bf 60}, 2343 (1988).

\bibitem{dupl-sal2}
B. Duplantier and H. Saleur, Phys. Rev. Lett. {\bf 59}, 539 (1987).

\bibitem{meiro} 
I. Chang and H. Meirovitch, Phys. Rev. {\bf E 48}, 3656 (1993).

\bibitem{hegg-grass} 
P. Grassberger and R. Hegger, Journal de Physique I 5, 597 (1995).

\end{thebibliography}
\end{document}